\documentclass[12pt]{article}
\usepackage{amsmath}
\usepackage{amssymb}

\begin{document}
\begin{center}
{\Large \textbf{A family of heavenly metrics}}\\[4mm] {\large\textbf{Y.
Nutku}} (deceased) and {\large\textbf{M. B. Sheftel$^1$}}
\\[2mm]
$^1$ Bo\u{g}azi\c{c}i University 34342 Bebek, Istanbul, Turkey
\end{center}
\vspace{1mm}
{\small\textit{\indent This is a corrected and essentially extended version
of the previous preprint gr-qc/0105088 v4 (2002) by Y. Nutku
and M. Sheftel which contains new results. It is proposed to be published in honor of Y. Nutku's memory.
All corrections and new results are due to M. Sheftel and concern only sections \ref{sec:intro},
\ref{sec:heavmetric} and \ref{sec:conclu}.}}
\vspace{3mm}\\
{\bf Abstract:} We present new anti-self-dual exact solutions of the Einstein
field equations with Euclidean and neutral (ultra-hyperbolic) signatures that
admit only one rotational Killing vector. Such solutions of the Einstein field equations
are determined by non-invariant solutions of Boyer-Finley ($BF$) equation.
For the case of Euclidean signature such a solution of the $BF$ equation
was first constructed by Calderbank and Tod.
Two years later, Martina, Sheftel and Winternitz applied the method of group
foliation to the Boyer-Finley equation and reproduced the
Calderbank-Tod solution together with new solutions for the neutral signature.
In the case of Euclidean signature we obtain new metrics
which asymptotically locally look like a flat space and have a non-removable singular point at the origin.
In the case of ultra-hyperbolic signature there exist three
inequivalent forms of metric. Only one of these can be obtained by
analytic continuation from the Calderbank-Tod solution whereas the
other two are new.


\begin{center}
\end{center}

\section{Introduction}
\label{sec:intro}

In his pioneer paper, Pleba\~nski showed that
Einstein field equations with Euclidean signature for anti-self-dual
($ASD$) metrics may be reduced to complex elliptic Monge-Amp\`ere equation
($CMA$), a real cross section of his first heavenly equation \cite{pleb}.
For elliptic $CMA$, solutions determine hyper-K\"ahler
$ASD$ Ricci-flat metrics with Euclidean
signature. We shall consider a symmetry reduction of $CMA$ to
solutions that generically admit only one-parameter continuous symmetry group
and hence generate metrics with only a single Killing vector. Systematic
studies of vacuum metrics with one Killing vector \cite{fp,bf}
showed that they fall into two classes depending on whether the Killing vector is
translational, or rotational. In the first case we have
Gibbons-Hawking metrics \cite{gibh} where the Einstein field
equations reduce to three-dimensional Laplace equation, whereas
in the latter case the reduction is to the Boyer-Finley ($BF$) equation \cite{fp,bf},
also known as $SU(\infty)$ Toda equation \cite{toda,mart} and as $sDi\!f\!f(2)$ Toda equation \cite{Takasaki,Finley},
for which until recently no non-trivial exact solutions were known.
Here following Boyer and Finley \cite{bf},
we call the solutions to be non-trivial if they do not satisfy simultaneously the Laplace equation.
In both cases the Riemannian curvature $2$-forms are
anti-self-dual \cite{bf,gh0,gr,tw}. We refer to \cite{klnt} and
\cite{bfp} for the ${\cal H}$-space context of these solutions.

Until recently, all solutions of the $BF$ equation discussed in the
literature consisted of invariant solutions, {\it i.e.}
solutions that were invariant with respect to symmetry
groups of the heavenly equation. These symmetries are carried
over into the metric in the form of extra Killing vectors. We
shall present metrics derived from non-invariant solutions of the
$BF$ equation which generically will not possess any new symmetries.
For the case of Euclidean
signature a non-invariant solution of the Boyer-Finley
equation was recently obtained by Calderbank and Tod \cite{tod}.
 Two years later,
Martina, Sheftel and Winternitz \cite{msw} applied the method of
group foliation to the Boyer-Finley equation and obtained new
non-invariant solutions, being not aware about the solution of Calderbank and Tod.
In the framework of the latter method, the Calderbank-Tod solution emerged naturally and we
had rigorously proved its non-invariance in generic case together
with a complete list of particular cases corresponding to invariant solutions.
We note that Tod \cite{Tod} and, quite recently, Dunajski et. al. \cite{dun} obtained
separable solutions to the Boyer-Finley equation
which reduce in essence to solutions of Liouville equation.
These solutions are invariant, admitting one-parameter symmetry subgroup
of the conformal group.

For the case of
ultra-hyperbolic (neutral) signature there exist three inequivalent classes
of non-invariant solutions. Only one of them can be obtained by analytic
continuation from the Calderbank-Tod solution and the other two
are new. Later, another class of non-invariant solutions was obtained by
hodograph transformation in \cite{ManAlo}.

In section \ref{sec:heavmetric} we shall briefly recall the
reduction of $CMA$ to the $BF$ equation. We construct new
families of $ASD$ Ricci-flat metrics which have only one singular point at the origin
and admit a single rotational Killing
vector generated by non-invariant solutions of the heavenly
equation. In the generic case, when these non-invariant solutions do not reduce to invariant ones,
the corresponding metrics seemingly do not belong to the Gibbons-Hawking class.
We show explicitly for one particular choice of parameters, when the solutions admit the extra symmetry
group $SU(2)$, that the metric does belong to the Gibbons-Hawking class with potential $V$ being a
harmonic function though not of a multi-center form. This $V$ is real-valued, though it is written with the use of complex numbers.
For this example and a more general one, based on a non-invariant solution of the $BF$ equation,
we have calculated  explicitly Riemann curvature two-forms and demonstrated that the curvature
is mainly concentrated in a bounded region of the Riemannian space,
similar to the behavior of gravitational instantons.
We also show that asymptotically our metrics look locally, but not globally,
as a flat metric on $\mathbb{R}^4$ in spherical polar coordinates. However, our metrics are not complete
since they have an isolated singular point at the origin which cannot be removed and therefore they cannot
be interpreted as gravitational instantons.

Metrics with ultra-hyperbolic (neutral) signature are
presented in section~\ref{sec:ultra}. Such metrics are important for anti-self-dual
conformal structures where the condition of anti-self-duality
is imposed not on the Riemann tensor but, more generally, on the conformal Weyl curvature tensor, so that
$ASD$ Riemann tensor becomes a special case. Four-dimensional $ASD$ conformal
structures are important in the theory of integrable systems which can be obtained
by their symmetry reductions \cite{dunaj}.

\section{Solutions with Euclidean signature}
\label{sec:heavmetric}

We consider a two dimensional complex manifold endowed with a
K\"ahler metric
\begin{equation}
d s^2 =  u_{i \bar k} \, d z^i d {\bar z}^k \label{kahler}
\end{equation}
with a metric potential $u(z^1,z^2,\bar z^1,\bar z^2)$ satisfying elliptic $CMA$ equation
\begin{equation}
u_{1\bar 1}u_{2\bar 2} - u_{1\bar 2}u_{2\bar 1}=1 \label{cma}
\end{equation}
so that we are dealing with a Ricci-flat metric and the Riemann
curvature two-form is anti-self-dual. This equation is a real
cross section of the first heavenly equation of Pleba\~nski \cite{pleb}. We shall be
interested in solutions of \eqref{cma} invariant under rotations in the complex $z^1$-plane
where
\begin{equation}
X = i\left(z^1\partial_{1}-{\bar z}^1\partial_{\bar 1}\right)
\label{rotgen}
\end{equation}
is the generator. If we introduce the polar radius $\rho$ and the
angle $\omega$ in the complex $z^1$-plane
\[ z^1 = \frac{1}{2}\,\rho e^{i\tau} \]
then the Killing vector \eqref{rotgen} becomes
$X=\partial_\tau$, the generator of rotations on the
$z^1$-plane. Henceforth we shall refer to $z^2$ as simply $z$.
Then the symmetry reduction of $CMA$ \eqref{cma} results in
\begin{equation}
u_{\xi\xi}u_{z\bar z}-u_{\xi z}u_{\xi\bar z}=e^\xi \label{rcma}
\end{equation}
where $\xi = 2\ln{(\rho/2)}$. But now the metric \eqref{kahler}
assumes a complicated form
$$ \begin{array}{rcl}
d s^2 & = & u_{z\bar z}dz d\bar z
+u_{\xi\xi}\left(\frac{1}{4}d\xi^2
+d\tau^2\right)+\frac{1}{2}\,u_{\xi z}d\xi dz \nonumber \\
 & & +\frac{1}{2}\,u_{\xi\bar z}d\xi d\bar z-iu_{\xi z}d\tau dz
+iu_{\xi\bar z}d\tau d\bar z \end{array} $$ and in order to
simplify it further we apply the Legendre transformation
\begin{equation}
\xi = -\varphi_p(p,z,\bar z),\quad p=u_\xi(\xi,z,\bar z),\quad u =
\varphi - p\varphi_p \label{legendre}
\end{equation}
whereby the reduced equation determining rotationally invariant
solutions of $CMA$ becomes the potential Boyer-Finley equation $
\varphi_{z \bar z}+\varphi_{pp}e^{-\varphi_p} = 0, $ or by
introducing $$ w(p,z,\bar z)=-\varphi_p$$ we obtain
\begin{equation}
w_{z\bar z}+\left(e^w\right)_{pp} = 0 \label{heaven+}
\end{equation}
the $BF$ equation.  This equation is also known as
 $SU(\infty)$ Toda equation \cite{toda,mart} and as $sDi\!f\!f(2)$ Toda equation \cite{Takasaki,Finley}.
After the Legendre transformation \eqref{legendre} the K\"ahler
metric becomes
\begin{eqnarray}
d s^2 & = &  w_p \, d l^2 +\frac{1}{w_p}\left[2 d \tau + i( w_z \,
d z - w_{\bar z} \, d \bar z )\right]^2 \label{heavmetr+} \\
d l^2 & = & 4 e^w d z \, d \bar z + d p^2. \label{3metric}
\end{eqnarray}
Thus, the Boyer-Finley equation \eqref{heaven+}
determines anti-self-dual rotationally invariant solutions \eqref{heavmetr+}
of the Einstein equations with Euclidean signature.

Our metric \eqref{heavmetr+} at a first glance looks similar to
the Gibbons-Hawking metrics \cite{gh0}. However, we have found that the curvature tensor
of the spatial metric \eqref{3metric} does not vanish, apart from a trivial case
of constant real $a$, when the four-dimensional metric \eqref{heavmetr+} describes a flat space.
Thus, in non-trivial cases $d l^2$ is not flat, unlike the
Gibbons-Hawking class. Further difference between these two classes
of metrics is that in the Gibbons-Hawking case the factor $V$ before $d l^2$
satisfies three-dimensional Laplace equation whereas in our metric this
factor is $V = w_p$\,, where $w$ satisfies the $BF$ equation \eqref{heaven+}, which is not
a solution of the Laplace equation.

We shall now use our non-invariant solutions of the Boyer-Finley equation
\cite{msw} given by
\begin{equation}
w = \ln \left| \frac{\{ p+a(z) \} b'(z) }{ 1 + |\,b(z)|^2}
\right|^2 \label{sol+}
\end{equation}
where $a$ and $b$ are arbitrary holomorphic functions, one of
which can be removed by a conformal symmetry transformation.
Accordingly we can specialize either $a$, or $b$, or some
combination of these holomorphic functions to be $z$. This is a
matter that should be decided depending on the physical problem at
hand. Then we obtain one representative for each orbit of these
solutions with respect to the conformal group and the choice of
the remaining arbitrary holomorphic function specifies the
particular orbit. This solution was obtained by the method of
group foliation \cite{msw} but it was sufficiently simple to be
derived readily by Calderbank and Tod, earlier in \cite{tod}.
The gravitational metric resulting from
\eqref{sol+} is given by
\begin{eqnarray}
d s^2 & = & |2p + a +\bar  a| \left\{ d S_2^2 +\frac{1}{|p+a|^2}
\, d p^2 \right\}  \label{metrsol+} \\ & &  +
\frac{|p+a|^2}{|2p+a+\bar a|} \left\{2 (d\tau + A_M)  +  i \left(
\frac{a' d z}{p+a} -\frac{\bar a' d \bar z}{p+\bar a}\right)
\right\}^2 \nonumber
\end{eqnarray}
with
\begin{eqnarray}
 d S_2^2 & = &  \frac{4 \,|b'|^2}{(1 + |b|^2)^2 } \; d z \, d \bar z \label{sphere} \\
  A_M  & = & - i \left[ \left( \frac{ \bar{b} \, b^{\prime}}{1 +|b|^2}
  - \frac{b^{\prime \prime}}{2 b^{\prime}} \right) d z -
   \left( \frac{b \,  \bar{b}^{\prime}}{1 +|b|^2}
- \frac{\bar{b}^{\prime \prime}}{2 \bar{b}^{\prime}}  \right) d
\bar{z}  \right] \label{monopole}
\end{eqnarray}
where primes denote derivatives of functions of a single variable.
Here we introduced absolute value sign in $|2p + a +\bar  a|$ because the change of the sign of $2p+a+\bar a$
corresponds to the change of inessential overall sign in the metric \eqref{metrsol+}.
We have
already pointed out that there is only one essential arbitrary
holomorphic function here. Furthermore, the choice of the
remaining holomorphic function, that will pick out interesting
solutions in the family \eqref{sol+}, is not completely arbitrary.
For some holomorphic functions the solution of the heavenly
equation will reduce to an invariant solution which will introduce
extra Killing vectors in the metric. The complete list of possible
choices for the arbitrary functions that result in invariant
solutions can be found in \cite{msw}.

For purposes of physical interpretation the choice $b = z$ is an
interesting case. First of all, in the generic case of an
arbitrary holomorphic function $a$ the solution will be
non-invariant. Then expressing the independent complex variable by
the stereographic projection
\begin{equation}
z = e^{i \phi} \cot \frac{\theta}{2}
 \label{stereo}
\end{equation}
\eqref{sphere} and \eqref{monopole} become
\begin{eqnarray}
 d S_2^2 & = & d \theta^2 + \sin^2 \theta \, d \phi^2 \nonumber \\
 A_M  & = & ( 1 + \cos \theta ) \, d \phi
 \label{dS_2^2 A_M}
\end{eqnarray}
the metric on the $2$-sphere and the magnetic monopole potential
one-form. We point out that the transformation \eqref{stereo} is not defined at $\theta\ne 0$,
so that $\theta = 0$ presents a coordinate singularity and allowed range of $\theta$ is
$0<\theta\leqslant\pi$.

For the simplest choice of constant real $a$ in this category,
the Riemann curvature tensor for
the metric \eqref{metrsol+}
vanishes and therefore this solution yields a flat space.
The simplest nontrivial case that yields a curved space corresponds to $a$ being a complex constant
with the imaginary part $\alpha = \Im a\ne 0$.
The metric \eqref{metrsol+} becomes
\begin{equation}\label{metrspec}
  d s^2 = \frac{4r^4 dr^2}{r^4+\alpha^2} + r^2 dS_2^2 + \frac{r^4+\alpha^2}{r^2}\{d \tau + A_M\}^2
\end{equation}
where we have canceled the overall factor 2, $r$ is radial coordinate, $r^2=|p+(a+\bar a)/2|=|p+\Re a|$,
$dS_2^2$ and $A_M$ are defined in \eqref{dS_2^2 A_M}.
This metric looks like an analytical continuation of the Eguchi-Hanson metric \cite{eghan,egh}
\begin{equation}\label{eguchi}
  d s^2 = \frac{r^4 dr^2}{r^4 - a^4} + \frac{r^2}{4} dS_2^2 + \frac{r^4 - a^4}{4r^2}\{d \psi + \cos{\theta}d\phi\}^2
\end{equation}
in the parameter $a^4$, notably replacing $a^4$ by $-\alpha^2$. The metric \eqref{eguchi} is known to be a two-center gravitational
instanton of the Gibbons-Hawking class \cite{gibh}
\begin{equation}\label{gibhaw}
    d s^2 = V(d\vec{R}\cdot d\vec{R}) + V^{-1}\{2d\phi + (\cos{\theta_+} + \cos{\theta_-})d\psi\}^2
\end{equation}
where $V$ is a particular two-center solution of the three dimensional Laplace equation
\begin{equation}\label{V}
    V = \frac{1}{r_+} + \frac{1}{r_-},\qquad r_\pm = \left|\vec{R}\pm \frac{a^2}{8}\hat{k}\right|,\quad \hat{k} = (0,0,1).
\end{equation}
Here $\vec{R}$ is a position vector from the origin and $r_\pm$ and $\theta_\pm$ are spherical polar coordinates with the origin at each of the two centers.
The Gibbons-Hawking metric \eqref{gibhaw} can be shown to coincide with the Eguchi-Hanson metric \eqref{eguchi} in the parametrization
\cite{Page,dunaj}
\[r_\pm = \frac{1}{8}(r^2 \pm a^2\cos{\theta}),\quad r_\pm \cos{\theta_\pm} =  \frac{1}{8}(r^2\cos{\theta} \pm a^2),\quad
V = \frac{16r^2}{r^4 - a^4\cos^2{\theta}}.\]
In order to try applying this to our metric \eqref{metrspec}, the replacement of $a^4$ by $-\alpha^2$ should be made in all the formulas
leading from \eqref{gibhaw} to \eqref{eguchi}. This will result in $V = 16r^2/(r^4 + a^4\cos^2{\theta})$ which cannot be expanded as in \eqref{V}
unless the spherical radii and angles become complex, $r_\pm = \frac{1}{8}(r^2 \pm i a^2\cos{\theta})$,
$r_\pm \cos{\theta_\pm} =  \frac{1}{8}(r^2\cos{\theta} \pm i a^2)$, which of course cannot be allowed. The function $V$ in the
coordinate system centered at the origin has the form
\begin{equation}\label{V(R)}
    V = \frac{\sqrt{R^2-a^4/64+i(a^2/4)R\cos{\vartheta}} + \sqrt{R^2-a^4/64-i(a^2/4)R\cos{\vartheta}}}{\sqrt{(R^2-a^4/64)^2 + (a^4/16)R^2\cos^2{\vartheta}}}
\end{equation}
where $R$ and $\vartheta$ are spherical polar coordinates of $\vec{R}$. The function $V$ in \eqref{V(R)} satisfies the three dimensional Laplace equation
in the coordinates of $\vec{R}$ and it is obviously real-valued though it is written with the use of complex numbers.
Thus, our metric, in the simplest nontrivial case \eqref{metrspec}, belongs to the Gibbons-Hawking class although it does not have the multi-center form
over the field of real numbers.

For our choice of $b=z$ and constant $a$, our solution of the $BF$ equation becomes $w = \ln{(p+a)} + \ln{(p-a)} - 2\ln{(1+|z|^2)}$,
which admits the obvious symmetry of rotations in the complex $z$-plane (originally $z^2$-plane) by the phase angle of $z$. Symmetry analysis
of the latter solution discovers two more non-obvious symmetries, so we have three symmetry generators for our solution and the metric
\begin{equation}\label{sym}
    X_1 = z^2\partial_z + \partial_{\bar{z}} - 2z\partial_w,\quad \bar{X}_1 = \bar{z}^2\partial_{\bar{z}} + \partial_z - 2\bar{z}\partial_w,
    \quad X_2 = z\partial_z - \bar{z}\partial_{\bar z}
\end{equation}
which satisfy the $SU(2)$ commutator relations
\begin{equation}\label{commut}
   [X_2,X_1] = X_1,\quad [X_2,\bar{X}_1] = - \bar{X}_1,\quad [X_1,\bar{X}_1] = - 2X_2.
\end{equation}
One more Killing vector $X_0 = \partial_\tau$ comes from the $BF$ equation being a rotational reduction of $CMA$.
Therefore, the metric \eqref{metrspec} in our simple example admits the symmetry group $U(1)\times SU(2)$.
We note that a general analysis of solutions of the $BF$ equation with $SU(2)$ symmetry was recently performed in \cite{Finley}
with the full set of such solutions obtained in a parametric form, though no explicit solutions and the metric were presented.

To compute Riemann curvature two-forms, we choose the tetrad coframe to be
\begin{eqnarray}
   & & e^0 = \frac{2r^2\,d r}{\sqrt{r^4+\alpha^2}},\quad e^1 = r d\theta ,\quad e^2 = r\sin{\theta}d \phi \nonumber \\
   & & e^3 = \frac{\sqrt{r^4+\alpha^2}}{r}\,\{d\tau + (1 + \cos{\theta})d\phi\}
   \label{tetrad}
\end{eqnarray}
so that the metric \eqref{metrspec} takes the diagonal form
\[d s^2 = (e^0)^2 + (e^1)^2 + (e^2)^2 + (e^3)^2.\]
With respect to this tetrad, running the package EXCALC under REDUCE \cite{excalc},
we obtain the following non-zero Riemann curvature two-forms
\begin{eqnarray}
   & & R^0_{\ 1} = - R^1_{\ 0} = R^2_{\ 3} = - R^3_{\ 2} = \frac{\alpha^2}{2r^6}\,(e^0\wedge e^1 + e^2\wedge e^3) \nonumber \\
   & & R^0_{\ 2} = - R^2_{\ 0} = R^3_{\ 1} = - R^1_{\ 3} = \frac{\alpha^2}{2r^6}\,(e^0\wedge e^2 - e^1\wedge e^3) \nonumber \\
   & & R^3_{\ 0} = - R^0_{\ 3} = R^2_{\ 1} = - R^1_{\ 2} = \frac{\alpha^2}{r^6}\,(e^0\wedge e^3 + e^1\wedge e^2)
   \label{curv}
\end{eqnarray}
and check that the curvature is indeed anti-self-dual. From \eqref{curv} we see immediately that
the choice $\alpha=0$, i.e. $\bar a = a$, indeed yields a flat space,
so that the simplest choice of the parameter $a$, which corresponds to a curved space, is $\bar a = - a\ne 0$ with $r^2=|p|$.
In particular, the simplest possible choice is $a=i$, $\bar a = -i$ with $\alpha=1$. Changing the real part of $a$ will not
affect the metric \eqref{metrspec} and curvature tensor as is obvious from \eqref{curv}.

We observe that all the Riemann curvature two-forms fall off rapidly when the radius $r$ goes to infinity,
so the curvature is mostly concentrated in a finite region of the Riemannian space-time.
The existence of an isolated singular point $r=0$ implies the restriction $r>0$, similar to the restriction $r>a$
for the Eguchi-Hanson metric \eqref{eguchi}, but as opposed to the latter case the singularity at $r=0$ is not removable.

Asymptotically at large $r$ our metric \eqref{metrspec} becomes
\begin{eqnarray}
   & & ds^2 \sim 4dr^2 + r^2\left\{d\theta^2 + \sin^2{\theta} d\phi^2 + (d\psi + \cos{\theta} d\phi)^2\right\} \nonumber
\\ & & = 4dr^2 + r^2\left(\sigma_1^2 + \sigma_2^2 + \sigma_3^2\right)
\label{asympt}
\end{eqnarray}
where we have denoted $\psi = \phi + \tau$ and $\sigma_i$ are the standard left-invariant one-forms on the group manifold $SU(2)$
defined by \cite{dunaj,egh}
\[\sigma_1 + i\sigma_2 = e^{-i\psi}(d\theta + i\sin{\theta} d\phi),\quad \sigma_3 = d\psi + \cos{\theta} d\phi .\]
This metric would look like a flat metric if not for the range of the angle $\psi$.
Indeed, $0\leqslant\phi < 2\pi$ and $0\leqslant\tau < 2\pi$, the latter being an angle of rotation in the complex $z^1$-plane.
Therefore, the range of $\psi$ depends on the value of $\phi$ at each fixed $\phi$: $\phi\leqslant\psi<\phi + 2\pi$
and so the metric \eqref{asympt} is not flat.

Next we provide a simplest possible example of a gravitational metric which is generated by a non-invariant
solution of the Boyer-Finley equation and hence admits only a single Killing vector, inherited from the derivation of the $BF$
equation out of $CMA$ by the rotational symmetry reduction. For this purpose, we choose
\begin{equation}\label{a_z}
    a = z = e^{i \phi} \cot \frac{\theta}{2}
 \end{equation}
where we have used the stereographic projection \eqref{stereo}. The resulting solution of the $BF$ equation
\begin{eqnarray}
 w &=& \ln{(p^2 + p(z + \bar z) + |z|^2)} - 2\ln{(1 + |z|^2)} \nonumber \\
   &=& \ln{(r^4 + \cot^2{(\theta/2)}\sin^2{\phi})} - 2\ln{(1 + \cot^2{(\theta/2)})}
  \label{sol}
\end{eqnarray}
where $r^2 = p + (z+\bar z)/2 = p + \cos{\phi}\cot{(\theta/2)}$,
does not admit any symmetry of the $BF$ equation and therefore it is a non-invariant solution \cite{msw}.

The metric \eqref{metrsol+} becomes
\begin{equation}\label{metrnoninv}
    d s^2 = (e^0)^2 + (e^1)^2 + (e^2)^2 + (e^3)^2
\end{equation}
where the coframe tetrad is defined by
\begin{eqnarray}
   & & e^0 = \frac{r}{\sqrt{r^4 + \cot^2{\frac{\theta}{2}}\,\sin^2{\phi}}} \left(2r\,d r
   + \cot{\frac{\theta}{2}}\sin{\phi}\,d \phi + \frac{\cos{\phi\,d \theta}}{2\sin^2{\frac{\theta}{2}}}\right)
   \nonumber \\[2mm]
   & & e^1 = r d\theta ,\qquad e^2 = r\sin{\theta}d \phi
   \label{tetrad2}
  \\[2mm]
   & & e^3 = \frac{1}{r}\,\sqrt{r^4+\cot^2{\frac{\theta}{2}}\,\sin^2{\phi}}\,\left.\Bigg\{d\tau + (1 + \cos{\theta})d\phi\right. \nonumber \\
   & & \left.\mbox{} + \frac{(r^2-\cot{\frac{\theta}{2}}\cos{\phi})\sin{\phi}\,d \theta - (r^2\cos{\phi}
  + \cot^2{\frac{\theta}{2}}\,\sin^2{\phi})\sin{\theta}\,d \phi}{2\sin^2{\frac{\theta}{2}}\left(r^4+\cot^2{\frac{\theta}{2}}\,\sin^2{\phi}\right)}\right\}.
   \nonumber
\end{eqnarray}
The metric \eqref{metrnoninv} with the definitions \eqref{tetrad2} admits only a single rotational Killing vector
and it is seemingly not of the Gibbons-Hawking form.

We have computed all Riemannian curvature two-forms with respect to coframe \eqref{tetrad2} which are too lengthy for exhibiting all of them here.
As a typical example, we present here the curvature two-forms determined by the formulas of a least possible length
\begin{equation}\label{curv2}
    R^2_{\ 1} = - R^1_{\ 2} = - R^0_{\ 3} = \frac{F(r,\theta,\phi) - G(r,\theta,\phi)}{2r^6(1-\cos{\theta})^2
    \sqrt{\displaystyle r^4\sin^2{\frac{\displaystyle\theta}{\displaystyle 2}}
    +\cos^2{\frac{\displaystyle\theta}{\displaystyle 2}}\sin^2{\phi}}}
\end{equation}
where we have used the following shorthand notation
\begin{eqnarray}
   & & F = 2\sin{\frac{\theta}{2}}\left(r^4\sin^2{\frac{\theta}{2}}-3\cos^2{\frac{\theta}{2}}\sin^2{\phi}\right)
    \nonumber \\
   & & \times\left[\sin{\phi}\left(e^0\wedge e^1 + e^2\wedge e^3\right) - \cos{\phi}\left(e^0\wedge e^2 - e^1\wedge e^3\right)\right]  \nonumber \\
   & & - \left(1 - 2\sin^2{\theta}\sin^2{\phi}\right)\sqrt{r^4\sin^2{\frac{\theta}{2}} + \cos^2{\frac{\theta}{2}}\sin^2{\phi}}
   \left(e^0\wedge e^3 + e^1\wedge e^2\right), \nonumber \\
   & &  G = 4r^2\sin{\frac{\theta}{2}}\sin{\theta}\sin{\phi}\left[\cos{\phi}\left(e^0\wedge e^1 + e^2\wedge e^3\right)\right. \nonumber\\
   & & \left.\mbox{} \phantom{G =} + \sin{\phi}\left(e^0\wedge e^2 - e^1\wedge e^3\right)\right].
   \label{FG}
\end{eqnarray}
We see that the only singularities of the metric \eqref{metrnoninv} and curvature two-forms \eqref{curv2}, determined by \eqref{FG}, are poles at $r=0$ and $\theta=0$. Thus, we have again the only singular point at the origin of the spherical polar system which cannot be removed, whereas $\theta=0$ is the singularity of the coordinate transformation by the stereographic projection \eqref{stereo} which implies $\theta > 0$. The latter singularity is removable by locally changing the coordinate system to Cartesian system
back to complex coordinates $z, \bar z$ near $\theta=0$. The curvature two-forms \eqref{curv2} on account of \eqref{FG} decrease as $1/r^4$ as $r$ approaches infinity, with other curvature two-forms behaving similarly, so curvature is again mainly concentrated in a finite domain.

The full metric \eqref{metrsol+} has apparent curvature singularities at
\begin{equation}
p +  a = 0,\quad p + \bar a = 0,\quad 2p + a + \bar a = 0  \quad\textrm{and} \qquad b' = 0 .
\label{sing}
\end{equation}
In particular, for the choice $b=z$ the
singularity at $b^{\,\prime}=0$ is absent. The pole singularities at $p=a$ and $p=\bar a$
exist only if $a(z)$ and $\bar a(\bar z)$ are real-valued and hence $a = \bar a$ is a constant,
which is the trivial case of a flat space. The only isolated point singularity $2p + a + \bar a = 0$ at the origin $r=0$
is not removable.

\section{Metrics with ultra-hyperbolic signature}
\label{sec:ultra}

Euclidean metrics with self-dual curvature do not admit Lorentzian
sections and this is reflected by well-known obstacles that
$CMA$ presents to analytic continuation because it is the
equation that replaces Laplace's equation for functions of many
complex variables. This property is inherited by the Boyer-Finley
equation. We can however construct metrics with ultra-hyperbolic
signature and there are alternative ways of arriving at such
metrics starting with \eqref{heavmetr+}. The simplest choice is
\begin{equation}
d s^2 = w_p \left[ 4 e^w d z \, d \bar z - d p^2 \right]
 - \frac{1}{w_p}\left[ d t +  i ( w_z \, d z - w_{\bar z} \, d \bar
 z ) \right]^2 \label{heavmetr-}
\end{equation}
where the only Killing vector is a null boost instead of a rotation.
We shall present the alternative choices later. In this version we
keep $z$ as a complex coordinate. The Einstein field equations are
now reduced to
\begin{equation}
w_{z \bar z}-\left(e^w \right)_{pp} = 0 \label{heaven-}
\end{equation}
which is a hyperbolic version of the Boyer-Finley equation. It is the
reduced equation obtained by Legendre transformation from
hyperbolic $CMA$ which carries minus one on the right hand side
of \eqref{cma}.

The non-invariant solution of the hyperbolic $BF$ equation
\eqref{heaven-} analogous to \eqref{sol+} is given by  \cite{msw}
\begin{equation}
w =\ln \left| \frac{ \{ p+a(z) \} b'(z)}{b(z) + \bar{b}(\bar z)}
\right|^2 \label{sol-}
\end{equation}
and the resulting $ASD$ Ricci-flat metric with ultra-hyperbolic signature becomes
\begin{eqnarray}
d s_1^2 & = & (2p + a +\bar  a ) \left[ \frac{4 |b^\prime|^2}{(b +
\bar{b})^2 } \; d z \, d \bar z  - \frac{1}{|p+a|^2} \;d p^2
\right]
 \nonumber \\
& &  - \frac{|p+a|^2}{(2p+a+\bar a)} \left\{ d t +  i \left[
\left( \frac{2 \, \bar{b}^\prime}{b + \bar b}  -
\frac{\bar{b}^{\prime \prime}}{\bar{b}^{\prime}} -
\frac{\bar{a}^\prime}{ p +
\bar{a}} \right) d \bar{z} \right. \right. \label{metrsol-}  \\
& & \hspace{1cm}  \left. \left. - \left( \frac{2 \, b^\prime}{b +
\bar b} - \frac{b^{\prime \prime}}{b^{\prime}} - \frac{a^\prime}{
p + a} \right) d z  \right] \right\}^2 \nonumber
\end{eqnarray}
where once again one of these holomorphic functions can be removed
by a conformal transformation. We note that the $2$-sphere metric
\eqref{sphere} in \eqref{metrsol+} is now replaced by the metric
for the Poincar\'e upper half plane with constant negative
curvature in \eqref{metrsol-}. This is immediate from the choice $
b =z, \; a=a(z)$. The singularity structure of this solution is
precisely the same as in \eqref{sing}.

An alternative form of the solution with ultra-hyperbolic
signature is obtained by regarding $z, \bar z$ as real null
coordinates $u, v$. Then there are two further inequivalent
metrics. The first one is given by
\begin{eqnarray}
d s_2^2 & = & (2p + a + b ) \left[ \frac{4 f^\prime g^\prime
}{(f+g)^2 } \; d u \, d v  - \frac{1}{(p+a ) (p+b)} \;d p^2
\right]
 \nonumber \\
& &  + \frac{(p+a)(p+b)}{(2p+a+b)} \left\{ d t +  \left( \frac{2
\, g^\prime }{f + g}  - \frac{g^{\prime \prime}}{g^{\prime}} -
\frac{b^\prime}{ p +b } \right) d v \right. \label{metrsol-2}  \\
& & \hspace{1cm}  \left. - \left( \frac{2 \, f^\prime}{f + g} -
\frac{f^{\prime \prime}}{f^{\prime}} - \frac{a^\prime}{ p + a}
\right) d u  \right\}^2 \nonumber
\end{eqnarray}
where $a=a(u), b=b(v)$ and $ f=f(u), g=g(v)$ replacing the
holomorphic functions $a$ and $b$ respectively. Next we have
\begin{eqnarray}
d s_3^2 & = & (2p + a + b ) \left[ \frac{4 f^\prime g^\prime }{(1
+ f g)^2 } \; d u \, d v  + \frac{1}{(p+a ) (p+b)} \;d p^2 \right]
 \nonumber \\
& &  - \frac{(p+a)(p+b)}{(2p+a+b)} \left\{ d t +  \left( \frac{2
\,f \, g^\prime }{1 + f g }  - \frac{g^{\prime
\prime}}{g^{\prime}} -
\frac{b^\prime}{ p +b } \right) d v \right. \label{metrsol-3}  \\
& & \hspace{1cm}  \left. - \left( \frac{2 \,g \, f^\prime}{ 1 + f
g} - \frac{f^{\prime \prime}}{f^{\prime}} - \frac{a^\prime}{ p +
a} \right) d u  \right\}^2 \nonumber
\end{eqnarray}
where again $a=a(u), b=b(v)$ and $ f=f(u), g=g(v)$. In these cases
the curvature singularities are at
\begin{equation}
p + a = 0, \quad p + b = 0,\quad 2p + a + \bar a =0, \qquad f^\prime = 0, \qquad g^\prime
= 0
\end{equation}
analogous to \eqref{sing}.

\section{Conclusion}
\label{sec:conclu}

We have discussed solutions of the Einstein field equations with
Euclidean or ultra-hyperbolic signature that generically admit only one
rotational Killing vector. They are determined by non-invariant
solutions of the elliptic or hyperbolic Boyer-Finley equation, respectively.
The first non-invariant solution of the elliptic Boyer-Finley equation
is due to Calderbank and Tod. It was obtained independently by
Martina, Sheftel and Winternitz by the method of the group foliation
of the Boyer-Finley equation together with full classification of particular cases
corresponding to invariant solutions, which imply extra Killing vectors in the metric.
By this method we have also constructed
new metrics with ultra-hyperbolic signature using our new
solutions of the hyperbolic version of the Boyer-Finley equation.
We found that there are three inequivalent classes of such
solutions, only one of which is manifestly derivable from the
Calderbank-Tod metric by analytic continuation.

In the case of Euclidean signature, we have constructed
the metric \eqref{metrsol+} which in the generic case admits only one
rotational Killing vector. It is a non-trivial example of
Boyer-Finley metrics because \eqref{sol+} is a non-invariant solution
of the Boyer-Finley equation which is not a simultaneous solution of
the Laplace equation.
We have presented a family of metrics \eqref{metrsol+}
with one arbitrary holomorphic function after a suitable
specialization of $a(z)$ and/or $b(z)$ is carried out, depending on the
physical problem at hand. For $b=z,\, \bar b= \bar z$ and constant $a$ and $\bar a$
(with the simplest nontrivial choice $\bar a= - a$) we have obtained
the anti-self-dual Ricci-flat metric \eqref{metrspec} for which the Riemann curvature tensor is
close to zero outside of a bounded domain. The metric \eqref{metrspec} belongs to the Gibbons-Hawking class
with the harmonic potential $V$ though it is not of a multi-center form.
This metric is incomplete because of the non-removable singular point at the origin.

For this particular simple choice of parameters $a$ and $b$ the metric in our
example, which we studied here in detail, admits three additional Killing vectors. These Killing vectors, together
with the rotational Killing vector used for the reduction of $CMA$ to $BF$ equation, generate Lie algebra
$U(1)\times SU(2)$. On the contrary,
our full metric \eqref{metrsol+} generically admits only a single rotational Killing vector.
To illustrate this, we present another example specializing $a$ as $a=z$ and $\bar a = \bar z$, which
results in another $ASD$ Ricci-flat metric \eqref{metrnoninv} determined by the coframe \eqref{tetrad2}.
This metric admits a single Killing vector. It has again the only non-removable singular point at the origin
and its curvature two-forms rapidly decrease at infinity. It does not seem to belong to the Gibbons-Hawking class.

It may be interesting to study different specializations of arbitrary functions
$a(z)$ and $b(z)$ which will correspond to $ASD$ Ricci-flat metrics of Euclidean signature admitting only one
rotational Killing vector.

\section{Acknowledgement}

We thank G. W. Gibbons for calling our attention to the paper of
Calderbank and Tod \cite{tod} in 2002. M. B. Sheftel thanks M. Dunajski
for illuminating discussions and comments and
A. A. Malykh for useful discussions.
The research of M. B. Sheftel was supported in part by the research grant
from Bo\u{g}azi\c{c}i University Scientific Research Fund (BAP),
project No. 6324.

\end{document}